\documentclass[aps,prd,twocolumn,superscriptaddress,showpacs]{revtex4}

\usepackage[dvips]{graphicx}
\usepackage{xspace}
\usepackage{latexsym}
\usepackage{amssymb}
\usepackage{times}
\usepackage{mathptm}

\begin{document}

\title
{A GEANT-based study of atmospheric neutrino oscillation 
parameters at INO}

\date{\today}

\newcommand{\sinp}{\affiliation{Saha Institute of Nuclear
Physics, 1/AF, Bidhannagar, Kolkata 700 064, India}}
\newcommand{\hri}{\affiliation{Harish-Chandra Research Institute, 
Chhatnag Road, Jhusi, Allahabad 211 019, India}}
\newcommand{\cu}{\affiliation{Department of Physics, University of
Calcutta,
92 Acharya Prafulla Chandra Road, Kolkata 700 009, India }}

\author {Abhijit Samanta}\sinp
\author {Sudeb Bhattacharya}\sinp
\author {Ambar Ghosal}\sinp 
\author {Kamales Kar}\sinp 
\author {Debasish Majumdar}\sinp
\author {Amitava Raychaudhuri}\hri\cu

\begin{abstract}
We have studied the dependence of the allowed space of the atmospheric neutrino 
oscillation parameters on the time of exposure for a magnetized Iron 
CALorimeter (ICAL) detector at the India-based Neutrino Observatory (INO). 
We have performed a Monte Carlo simulation for a 50 kTon ICAL detector generating 
events by the neutrino generator NUANCE and simulating the detector response 
by GEANT. A chi-square analysis for the ratio of the up-going and down-going 
neutrinos as a function of $L/E$ is performed and the allowed regions at
90\% and 99\% CL are displayed. { These results are found to be better
than the current experimental results of MINOS and Super-K.} The possibilities 
of further improvement  
have also been discussed.
\end{abstract}

\maketitle

\section{Introduction}
The evidence of neutrino masses and their mixing \cite{Fukuda:1998mi,Eidelman:2004wy}  has brought 
neutrino physics into centre stage of particle physics. The  
neutrino mass  eigenvalues and the Pontecorvo, Maki, Nakagawa, Sakata 
(PMNS) mixing matrix \cite{Pontecorvo:1957cp, Maki:1962mu} connecting the mass to the flavor basis 
provides  a natural framework for handling three active neutrinos. 

The present information on the neutrino mass-squared
differences and mixing angles are the following:
From atmospheric neutrino detection one gets the best-fit values with
$3\sigma$ error
$|\Delta m^{2}_{32}|\simeq 2.5^{+0.7}_{-0.6}\times 10^{-3}$
eV$^2$, $\sin^2\theta_{23}\simeq$ ${0.5}^{+0.18}_{-0.11}$
while solar neutrinos
tell us $\Delta m^{2}_{21} \simeq 7.9\times 10^{-5}$ eV$^2$,
$\sin^2\theta_{12}\simeq$ $0.30$ \cite{Schwetz:2006dh}.  
{Here we define $\Delta m^{2}_{ij}$= $m^{2}_{i} - m^{2}_{j}$.}

At the moment, the sign of $\Delta m^{2}_{32}$ is not known. 
The positive/negative value of this quantity denotes the  
direct/inverted mass ordering.  The two large mixing angles and the  
mass squared differences may permit measurement of CP-violation 
in the lepton sector, if the third mixing angle, $\theta_{13}$, and the CP
phase, $\delta$, are not too small.  The current bound on
the former is $\sin^{2}\theta_{13}$ $<$ 0.05 (3$\sigma$)
 \cite{Apollonio:1999ae,Bandyopadhyay:2004da} while $\delta$ is unconstrained.

Thus the determination of mass hierarchy and the measurement of oscillation
parameters with high precision are of utmost importance. { Also of importance
is observing a full oscillation cycle to convincingly  establish that it is
truly neutrino oscillation which is at play.
Most experiments observe the depletion part but not
the regeneration part of the cycle. A reanalysis of old Super-Kamiokande (Super-K) data claimed
to observe this \cite{Ashie:2004mr}. However, a reconfirmation of this 
with better statistics is much awaited. The mixing angle  and the mass squared
difference for the atmospheric sector should also be measured more
accurately. 
 }The sensitivity of 
the measurement of a particular parameter depends crucially on the ranges 
of neutrino energy and path length traversed  from the 
source to the detector. These ranges can be set in case of neutrino beams from 
artificial sources like nuclear reactors (energy $\sim$ MeV) and accelerators 
(energy $\sim$ GeV).  Neutrinos with energy $\sim$ MeV (GeV) can also be 
obtained from  natural sources like the sun (the atmosphere).
Unlike typical accelerator or reactor neutrinos, the spectrum of atmospheric
 neutrinos covers many decades of energy  ($E \sim$ 100 MeV --   
few hundred GeV) with comparable interaction rate and baseline
($L \sim$ 10 km - 12800 km). Since the oscillation probability depends
mainly on $L/E$ which varies in a wide range for atmospheric neutrinos, 
the measurement of the appearance/disappearance probability as a function of
$L/E$ can explore its variation over this entire range. This advantage is partly 
offset, however, by the difficulty  that the flux is less known compared 
to that from man-made sources. 

Currently around the world, there are many ongoing and planned experiments:
MINOS \cite{Zois:2004ns,Michael:2006rx}, T2K \cite{Yamada:2006hi}, ICARUS \cite{Kisiel:2005ti, Rubbia:1998rc},
NOvA \cite{Ray:2006ke, Harris:2005yb}, Double Chooz \cite{Horton-Smith:2006yh,
Motta:2006jd}, 
UNO \cite{Jung:1999jq}, Super-K III \cite{Back:2004qi}, 
Hyper-K \cite{Itow:2001ee, Nakamura:2003hk}, OPERA \cite{Cocco:2000yp, Gustavino:2006rc, Di Capua:2005bd} etc. Out of these
only MINOS employs a magnetic field and has a good charge identification 
capability. It is to be noted that all these experiments are planned in the
northern hemisphere of the earth.  

The proposed India-based Neutrino Observatory (INO) 
 \cite{Athar:2006yb} at a site close to the equator plans to use  
a large magnetized Iron CALorimeter (ICAL) detector. The proposal is
for an underground facility with more than 1 km overburden.
Since the detector has a high charge identification capability 
($>90\%$
after selection of events as described in section V  
and ~$70\%$ before doing the selection) \cite{Athar:2006yb},
it has a good chance of determining the
neutrino mass ordering \cite{Indumathi:2004kd, Gandhi:2004bj,Petcov:2005rv, Samanta:2006sj} 
and also of studying the deviation
from maximality for $\theta_{23}$ \cite{Choubey:2005zy, Indumathi:2006gr}.

In this work { we first demonstrate through a GEANT-based simulation
of atmospheric neutrinos that ICAL indeed is capable of observing 
the full oscillation cycle. We have used a two flavor oscillation formalism
and studied the precision that can be achieved for $|\Delta m^2|$ and 
$\sin^22\theta$ at INO with atmospheric neutrinos. Though a more
realistic approach would be the use of three flavor analysis,
the smallness of the mixing angle
$\theta_{13}$ ensures that the two flavor approximation mimics
the real situation reasonably well.}
The precision  depends on the 
exposure in terms of kTon-yr, reconstruction method, and the  selection
of the events in the analysis. The paper is organized as follows:
A brief summary of the neutrino oscillation formalism is given in Section II.
The ICAL detector at INO is
described in Section III. In Section IV a brief account of the
atmospheric neutrino flux that has been used in the present analysis
has been furnished. The generation of simulated data at ICAL and the
analysis of such data are described in Section V. In Section VI
we present the results and precision study of the oscillation parameters.
Finally,  Section VII includes discussions and conclusions.

\section{Neutrino oscillation}

A neutrino flavor eigenstate $|\nu_\alpha \rangle$
($\alpha \equiv e, \mu, \tau$ etc.) can be written
as a linear superposition of neutrino eigenstates $|\nu_i\rangle $
(with definite non-degenerate masses $m_i$)
in the mass basis as
$|\nu_\alpha \rangle = \sum_i U_{\alpha i} |\nu_i\rangle $ ($i = 1,2,3 $ etc.).
Here $U_{\alpha i}$ are the matrix elements of the neutrino mixing matrix $U$. 
This gives rise to the phenomenon of neutrino flavor oscillation.
The probability that a neutrino $\nu_g$
with energy $E$ gets converted into another neutrino $\nu_f$ 
 after traversing a distance $L$ in vacuum is given by
\begin{eqnarray}
P(\nu_{g} \rightarrow \nu_{f}) = \delta_{fg}
-4\sum_{j>i}\textrm{Re}(U^{\ast}_{fi}U_{gi}U_{fj}U^{\ast}_{gj})
\sin^{2}(1.27\Delta m^{2}_{ij}\frac{L}{E})\nonumber\\
\pm2\sum_{j>i}\textrm{Im}(U^{\ast}_{fi}U_{gi}U_{fj}U^{\ast}_{gj})
\sin(2.54\Delta m^{2}_{ij}\frac{L}{E})
\end{eqnarray}
In the above, $L$ is expressed in km, $E$ in GeV and $\Delta
m^{2}$ in eV$^{2}$. The ~ -- (+)~ refers to neutrinos (anti-neutrinos).

For a two flavor scenario the above equation takes a simplified form
given by
\begin{eqnarray}
P_{\rm survival}=1- \sin^22\theta\sin^2(1.27\Delta m^2 \frac{L}{E})
\end{eqnarray}
where $\theta$ is the mixing angle of neutrinos. 
Herein and in the rest of the paper the symbol $\theta$ and 
$\Delta m^2$ refer to  $\theta_{23}$ and $\Delta m_{32}^2$.

\section{The INO detector}

The simulation has been carried out for a detector with 50 kTon mass 
with dimension 
48 m $\times$ 16 m $\times$ 12 m for  
ICAL \cite{Athar:2006yb}. The detector consists of a 
stack of 140 horizontal layers of 6 cm thick iron slabs interleaved with 
2.5 cm gap for the active detector elements. For the sake of illustration,
we define a rectangular coordinate frame with origin at the center of the 
detector, $x(y)$-axis along the longest (shortest) lateral direction, and 
$z$-axis along the vertical direction. A magnetic field of strength 1 Tesla 
is considered along  the $+y$-direction. Resistive plate chambers (RPC) 
have been chosen as the active part of the detector. 
The readout of the RPCs is through the  Cu strips having 2 cm width and placed 
orthogonally on the two external sides of the detectors. This type of 
detector has good time ($\sim$ 1 ns) and spatial resolutions.

\section{Atmospheric neutrino flux}

The atmospheric neutrinos are produced from the interactions of the
cosmic rays with earth's atmosphere. The knowledge of the primary spectrum 
of cosmic rays has been  improved from the observations by 
BESS \cite{Sanuki:2000wh,Maeno:2000qx} and AMS \cite{Alcaraz:2000vp}. However, a  large region
of parameter space has been unexplored and they are interpolated 
or extrapolated from the measured flux.  The difficulties and uncertainties
in the calculation of the neutrino flux depend on the neutrino 
energy \cite{Honda:2004yz}.
The low energy flux is known quite well. The 
cosmic ray fluxes ($<$ 10 GeV)  are modulated by solar activity
and the geomagnetic field through a rigidity (momentum/charge) cutoff.
At higher neutrino energy ($>$ 100 GeV), solar activity and rigidity 
cutoff are irrelevant.  
There is an  agreement within  5\%  among the calculations for neutrino 
energy below 10 GeV though different groups used different hadronic 
interaction models in their calculations.

We use the neutrino interaction cross section model of NUANCE
 \cite{Casper:2002sd} incorporating a typical Honda flux calculated in 
a 3-dimensional scheme \cite{Honda:2004yz}. 
\section{Data generation and method of analysis}

The interactions of neutrinos with the detector material are simulated 
by the Monte Carlo method in NUANCE \cite{Casper:2002sd}. 
In order to study the ICAL detector response for each event we use  another Monte Carlo 
code GEANT \cite{geant}. The GEANT code uses the information of vertex position 
and  momentum of the product particles obtained from the output of NUANCE 
simulation.

{\bf Event reconstruction:} 
Our present analysis is based on the tracks generated by the muons that
are produced by the charged current (CC) interactions of the neutrinos in the detector volume. 
The muons lose energy mainly due to ionization and radiative processes.
This energy loss is proportional to the effective path-length which is 
the product of geometric path-length and the density of the medium. 
This can be applied only for fully contained (FC) events.

In this simulation study we do not consider any atmospheric muon background
and the noise hits produced by the detector. However, we conservatively 
assumed  hits $>$ 6 will be required for the reconstruction and filtering of the 
muon events from the latter background. 

For a given triggered energy of a muon,  the  number of hits  
decreases when one goes from vertical to horizontal direction since
it traverses less number of active detector elements.  
This  dependence on the direction is less for the effective path-length. 


In case of partially contained (PC) events  the momentum
has been determined from the curvature of the track at the vertex
due to the magnetic field applied across the detector. 
Due to  limitation of our PC event reconstruction algorithm
we considered  PC events with hits $<$ 20.

On the other hand, the hadrons produced create a shower of hits around 
the vertex of the event. {\it This implies that for any particular CC event  
the longest track normally comes  from  muons and this can be utilized for the 
analysis.} 


The up-going muon type neutrinos traverse larger path-length undergoing 
oscillation whereas the down-going ones with much shorter path-length have 
little chance to oscillate. 
So one can visualize our detector set up as  far (near) detector
for the up-going (down-going) neutrinos. 
Then the ratio of up-going and  down-going neutrinos (up/down) will  roughly 
mimic the survival probability. This up/down ratio as a function of $L/E$ 
minimizes the systematic uncertainties in flux as well as in cross sections. 
The length $L$ for the up-going neutrinos is the actual path-length
traversed by them whereas for down-going neutrinos the reference 
path-length $L$ is considered to be that of associated up-going neutrinos
with zenith angle ($180^\circ - \theta_{\rm zenith}$) so that the range 
of $L/E$ remains the same for up-going and down-going neutrinos
 \cite{mirrorL}.
 
\vspace*{.5cm}
{\bf Selection of events and Resolutions:}
\begin{figure}[thb]
\includegraphics[width=6.0cm,angle=270]{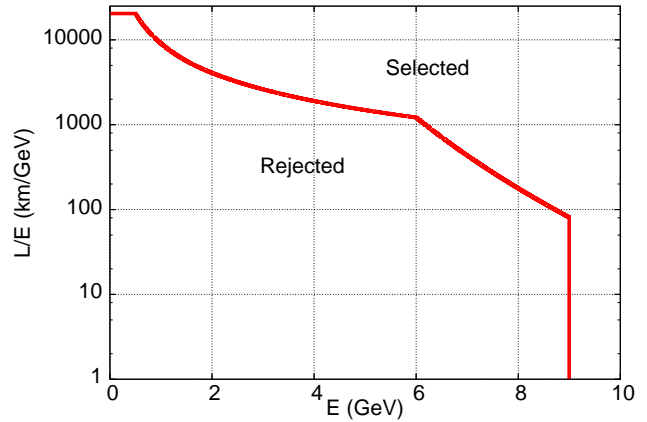}
\caption{\sf The selection of events in $E-L/E$ plane, which gives good optimization 
 between statistics and $L/E$ resolution.}
\label{f:cut}
\end{figure}

\begin{figure*}
\includegraphics[width=6.0cm,angle=270]{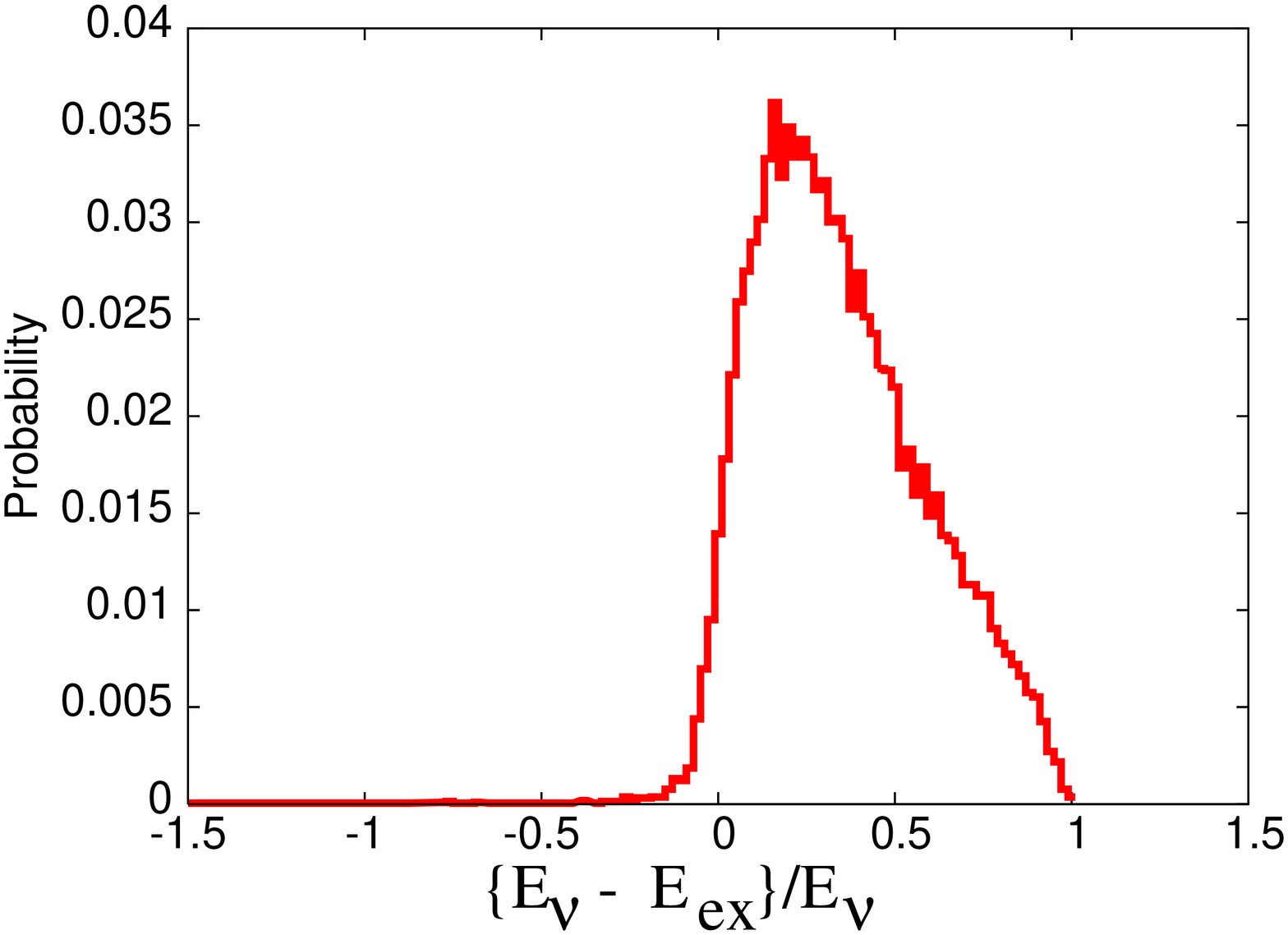}
\includegraphics[width=6.0cm,angle=270]{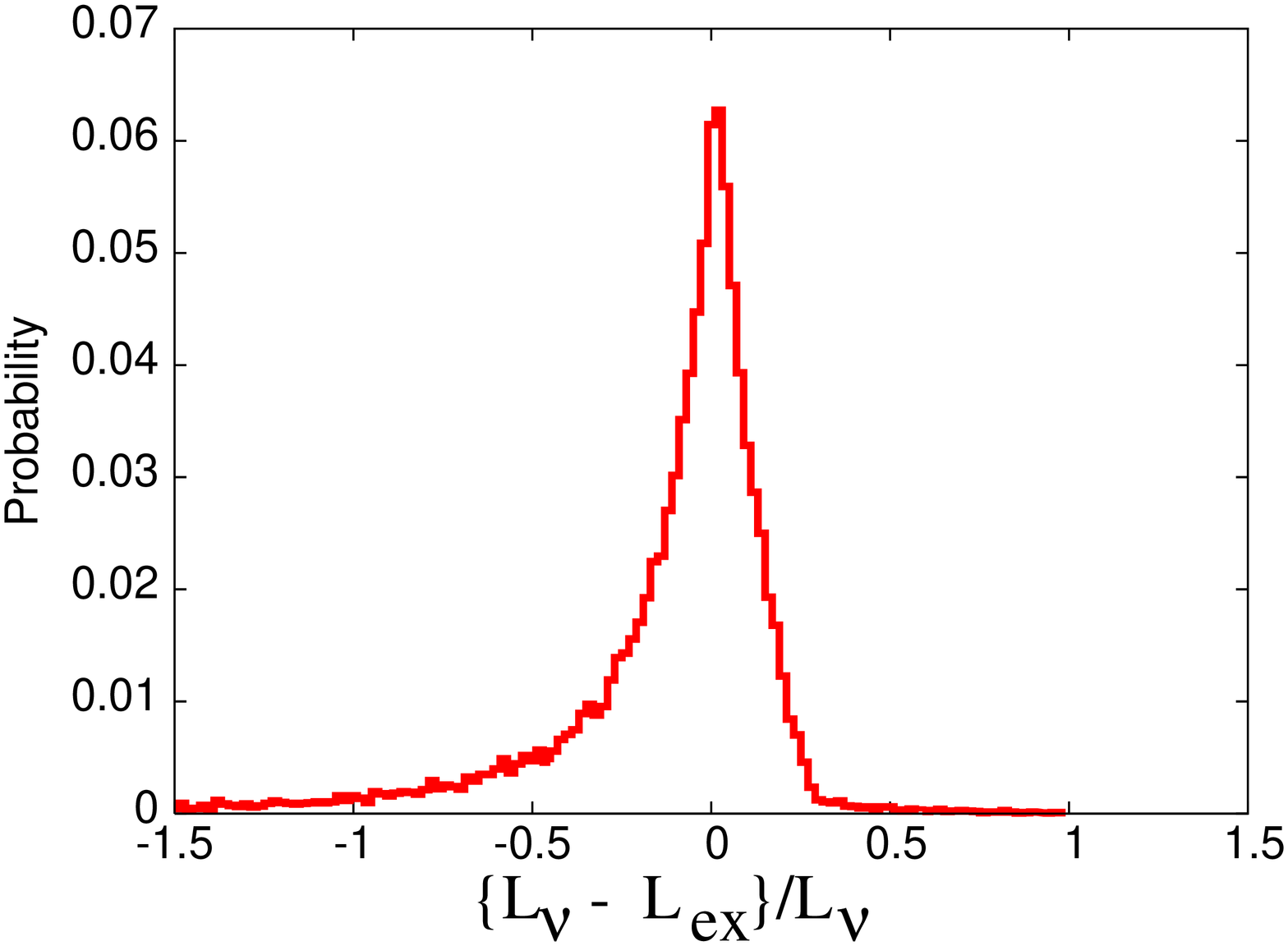}
\includegraphics[width=6.0cm,angle=270]{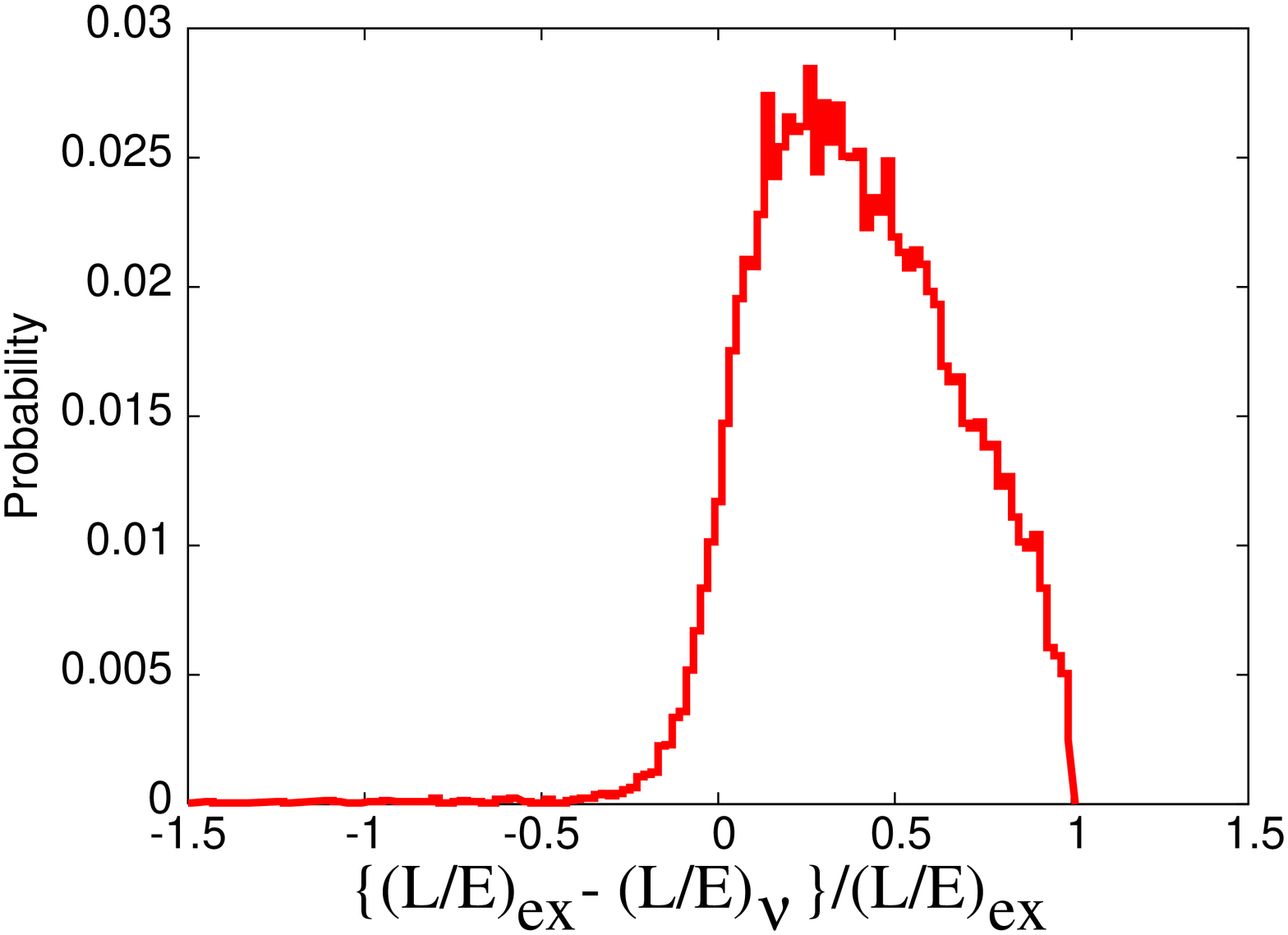}
\caption{\sf The ICAL resolutions for $E$, $L$ and $L/E$ with atmospheric 
neutrinos for the whole range of $E_\nu$ and $L_\nu$. The subscript `ex' 
and $\nu$ refer to the reconstructed and true $\nu$ values.}
\label{f:atmreso}
\end{figure*}


The $L/E$ resolution has a complicated dependence on $L$ and $E$. 
However, a few general remarks can be made here.
Qualitatively for a fixed energy, the $L$ resolution worsens gradually
as we go from vertical to horizontal region and worsens rapidly close to the
horizon. Also for a fixed direction, the $L/E$ resolution improves with increase 
in $E$. If one neglects totally the  near horizon events (say, between zenith
angle $70^\circ$ and $110^\circ$) all the events below 200 km/GeV are lost.
In our analysis we consider only the high energy events at near horizon
and relax it gradually as we move away from the horizon. Quantitatively,
this is taken care of by an $E$ dependent cut of the form :
$$ L/E \ge a E^b$$
broken into three segments as shown in fig. \ref{f:cut}.    

Using the above cut 
the resolutions for $E$, $L$ and $L/E$ obtained with the atmospheric neutrino flux 
for the whole range of $E_\nu$ and $L_\nu$ are shown in fig. \ref{f:atmreso}. 

A representative statistics for 5 year data  is shown in table \ref{t:statistics}.
Here the number of events with hits $>6$ is considered for the analysis. 
 The selected number of events is further  reduced by imposing the above $E$ dependent $L/E$ 
cut for a better resolution. 



\begin{table}[htb]
\begin{center}
\begin{tabular}{|c|c|c|c|c|}
\hline
 {\rm cut} & \multicolumn{4}{|c|}{No. of surviving events/efficiency} \\
\cline{2-5}& FC & efficiency & FC+PC& efficiency\\
\hline
{\rm hit$>$ 6}&4160& -&5351 &-\\
{\rm (for reconstruction)}&  & & &  \\
\hline
{\rm $E$ dependent $L/E$ cut} & {2089}& 50.2\% & 2808& {52.4\%}\\
{\rm (improves $L/E$ resolution)}& &&&  \\
\hline
\end{tabular}
\caption{\sf Sample number of events after  cuts in 5-year data for  
$\Delta m^2=2.3\times 10^{-3}$eV$^2$.}
\label{t:statistics}
\end{center}
\end{table}
\begin{figure*}[htb]
\includegraphics[width=6.0cm,angle=270]{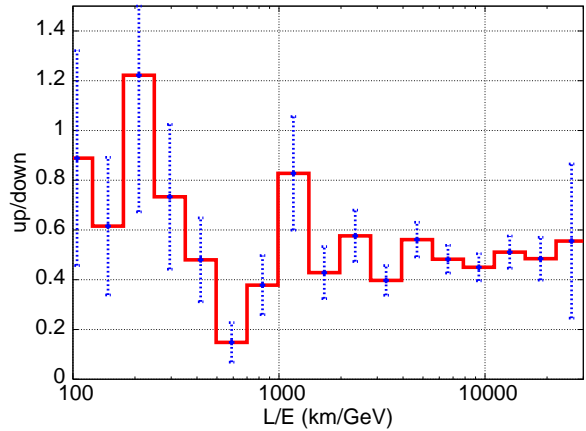}
\caption{\sf The simulated up/down distribution at ICAL  as a function of $L/E$ 
for 5 years FC events with $\Delta m^2=2.3\times 10^{-3}$eV$^2$. }
\label{f:ubd}
\end{figure*}

Using the above cuts we find the up/down distribution for
different time exposures of the ICAL detector. Here our main goal
is to find  how precisely one can  measure 
$\Delta m^2_{}$. 
A representative   `up/down' distribution with respect to $L/E$ for 5 year FC events 
is shown in fig. \ref{f:ubd}. Such simulated plots are referred to as the 
`experimental up/down' distributions. 

{\bf $\chi^2$-analysis:}

In the $\chi^2$-analysis the `theoretical up/down' distribution is obtained by 
taking 40 years of atmospheric un-oscillated charged 
current muon neutrino data. The oscillation probability  is then calculated 
from 
the  $L$ and $E$ of the neutrinos for each event and the event is kept or rejected  by throwing
a random number.
We smear this over the whole range of $L/E$ following the $L/E$-resolution function. 
Finally  the  up/down ratio is calculated for different 
$L/E$ bins. 
In this process we are also minimizing the effects due to  geomagnetism  
and the shape of the earth. 

Then a $\chi^2$-fit is made with the `experimental up/down' distribution
varying the atmospheric mass square difference $\Delta m^2$ and the 
mixing angle $\theta$ in the `theoretical up/down' data. 

\section{Results}

{ The $L/E$ plot in fig. \ref{f:ubd} shows clearly a full oscillation cycle 
and  is typical for the ICAL detector. Thus  ICAL is in a position to observe 
the 
oscillation pattern better than previous attempts, like the one by Super-K
 \cite{Ashie:2004mr}.    }

\begin{figure*}[htb]
\includegraphics[width=5.cm,angle=270]{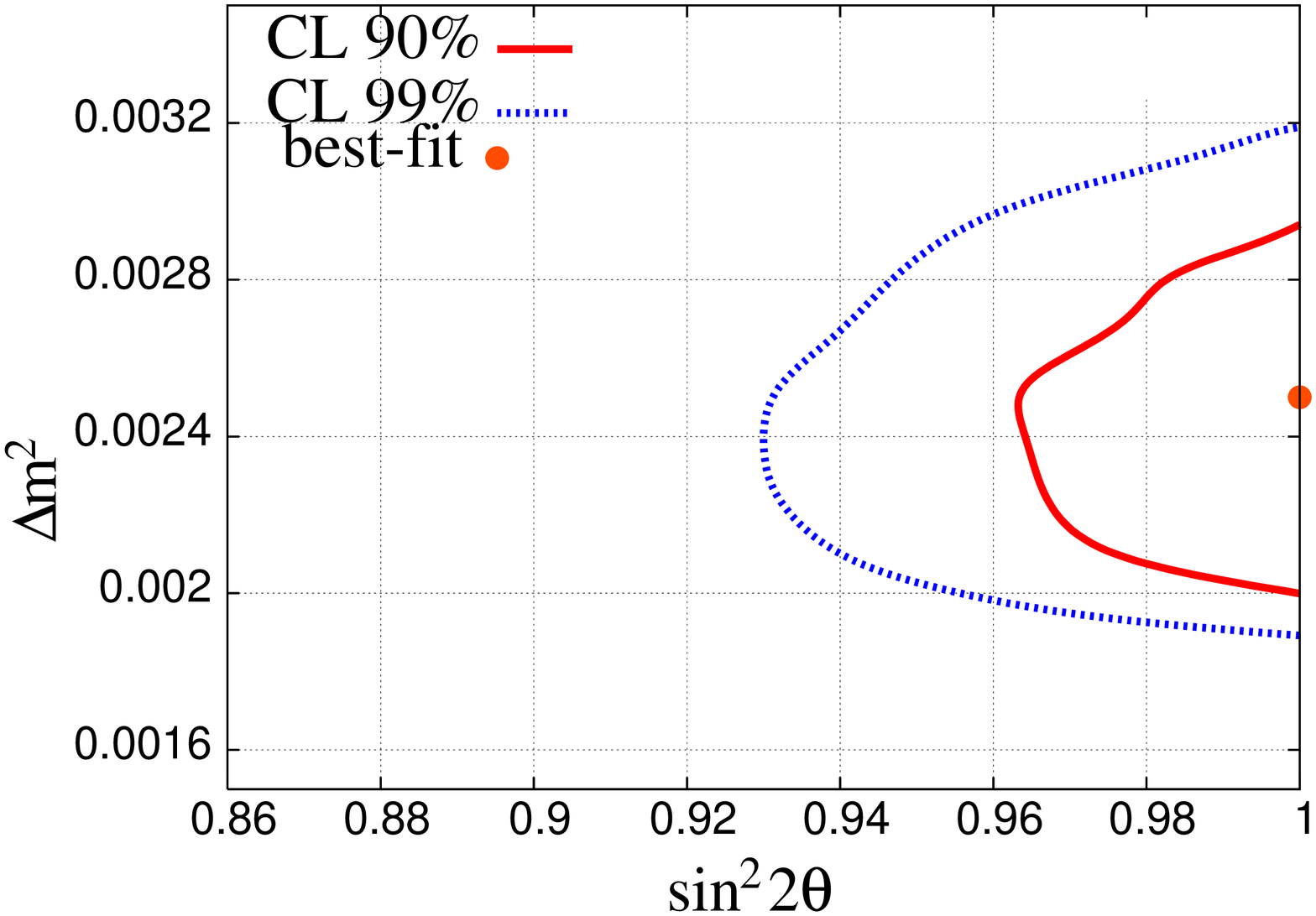}
\includegraphics[width=5.cm,angle=270]{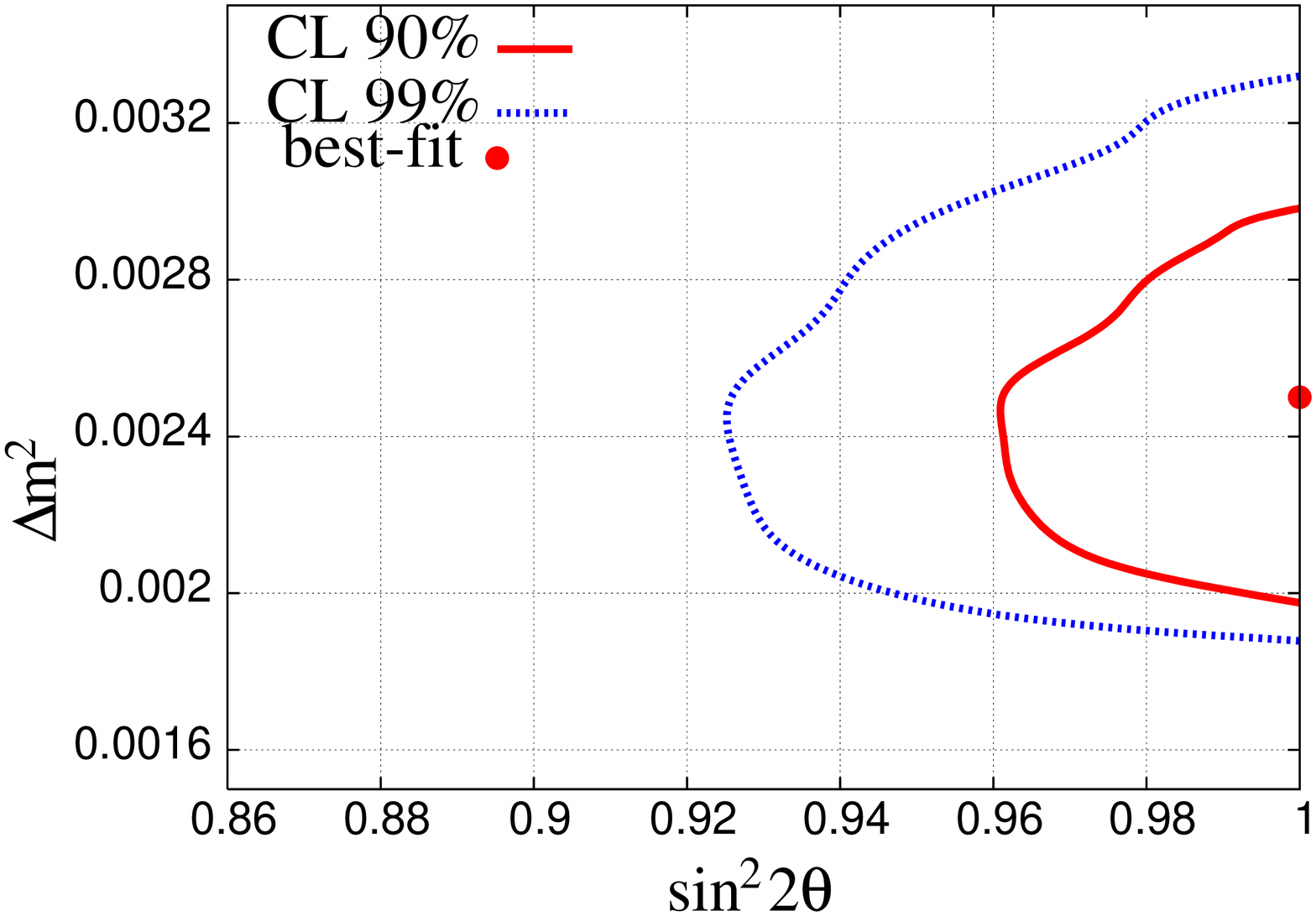}
\includegraphics[width=5.cm,angle=270]{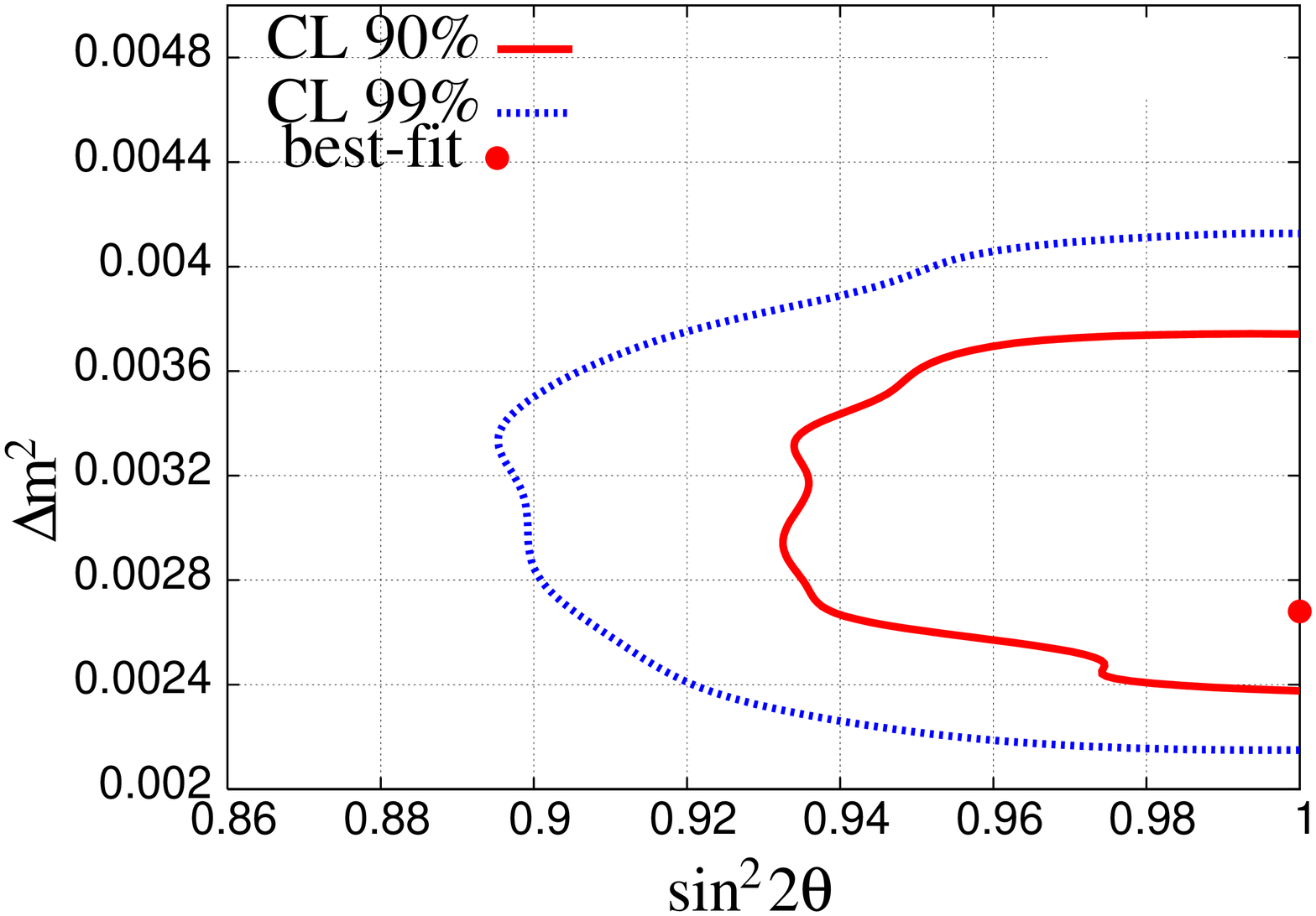}
\caption{\sf  The contour plots in $\Delta m_{}^2$ --
$\sin^22\theta_{}$ plane with input $\Delta m^2 = 2.3\times 10^{-3}$eV$^2$ for
 5 years FC events (upper left), FC+PC event (upper right); the lower plot is with 
input $\Delta m^2=2.7\times 10^{-3}$eV$^2$
for 10 years FC events.}
\label{f:contour}
\end{figure*}

We show in fig. \ref{f:contour} the contours in the $\Delta m_{}^2$ --- 
$\sin^22\theta_{}$ plane for 90\% and 99\% CL with  5 year FC  
(upper left) and 5 year FC+PC (upper right) events  for the input value of 
$\Delta m^2=2.5 \times 10^{-3}$eV$^2$
 and with 10 years FC events (lower) for $\Delta m^2=2.7 \times 10^{-3}$eV$^2$.

It is noted that the extracted best-fit values gradually become close 
to the input value of $\Delta m^2$ with increase of
exposure time. For example,  the 
best-fit value is found to be  2.50, 2.40 (2.50, 2.34) $\times 10^{-3}$eV$^2$ 
for 5, 10 years
FC (FC+PC) events with the input 2.3 $\times 10^{-3}$eV$^2$.
For all these cases the best-fit values of $\sin^22\theta$ turns out
to be 1 with the input value 1.  
Since the FC+PC sample contains more high energy events than the FC sample 
and the $E$ and $L$ resolution are better for PC events, the
best-fit values obtained from the FC+PC analysis are closer to the input value.
With a change of the input $\Delta m^2$ from 2.5 to 2.7 $\times$ 10$^{-3}$eV$^2$
for 10 year FC samples, the best-fit changes to 2.46 to 2.68 $\times$ 10$^{-3}$eV$^2$.



The position of the dip in this up/down distribution is indicative of the best-fit value 
of $\Delta m^2$  while  the overall statistics determines 
the size of the allowed parameter space. 
Particularly, the statistics in the larger (smaller) $L/E$ region
from the dip determines the lower (upper) bound of   $\Delta m^2$. 
For atmospheric neutrinos the statistics increases with increase of $L/E$
thus resulting in a better lower bound in the contour of $\Delta m^2$. 

\begin{figure*}[htb]
\includegraphics[width=6.0cm,angle=270]{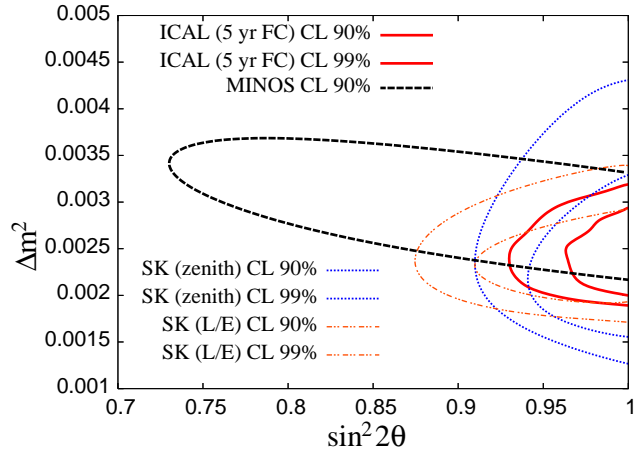}
\caption{\sf The contours at 90\% and 99\% CL 
for 5 years FC events with $\Delta m^2=2.3\times 10^{-3}$eV$^2$
at ICAL and the contours from the current experiments.}
\label{f:comp}
\end{figure*}

{ We have made a comparison of the present results and those 
obtained by the Super-K \cite{Ashie:2004mr,Ashie:2005ik} and the MINOS \cite{Michael:2006rx} experiments. They are shown in fig. 
\ref{f:comp}. Here we plot contours at 90\% and 99\% CL
in $\Delta m^2 - \sin^22\theta$ plane as obtained from our study 
of 5 years FC data and those obtained from previous Super-K (1489 days data) 
and the recent MINOS data. For the case of Super-K, two different analyses,
one with respect to zenith angle \cite{Ashie:2005ik} and the other with 
respect to $L/E$ \cite{Ashie:2004mr}
are given. One can see clearly that ICAL results are far more
precise than those of MINOS and substantially better than those from Super-K.} 

\subsection{Precision}

We define precision (${\rm P}$) for a certain 
confidence level of a particular set of oscillation parameters as  
$${\rm P} = 2 \,\left ( \frac{\rm UL - LL}{\rm UL + LL}\right )~$$
where `UL'  and `LL' are the upper and lower limit of the contour
respectively at the specified confidence level.

\begin{figure*}[htb]
  \includegraphics[width=6.cm,angle=270]{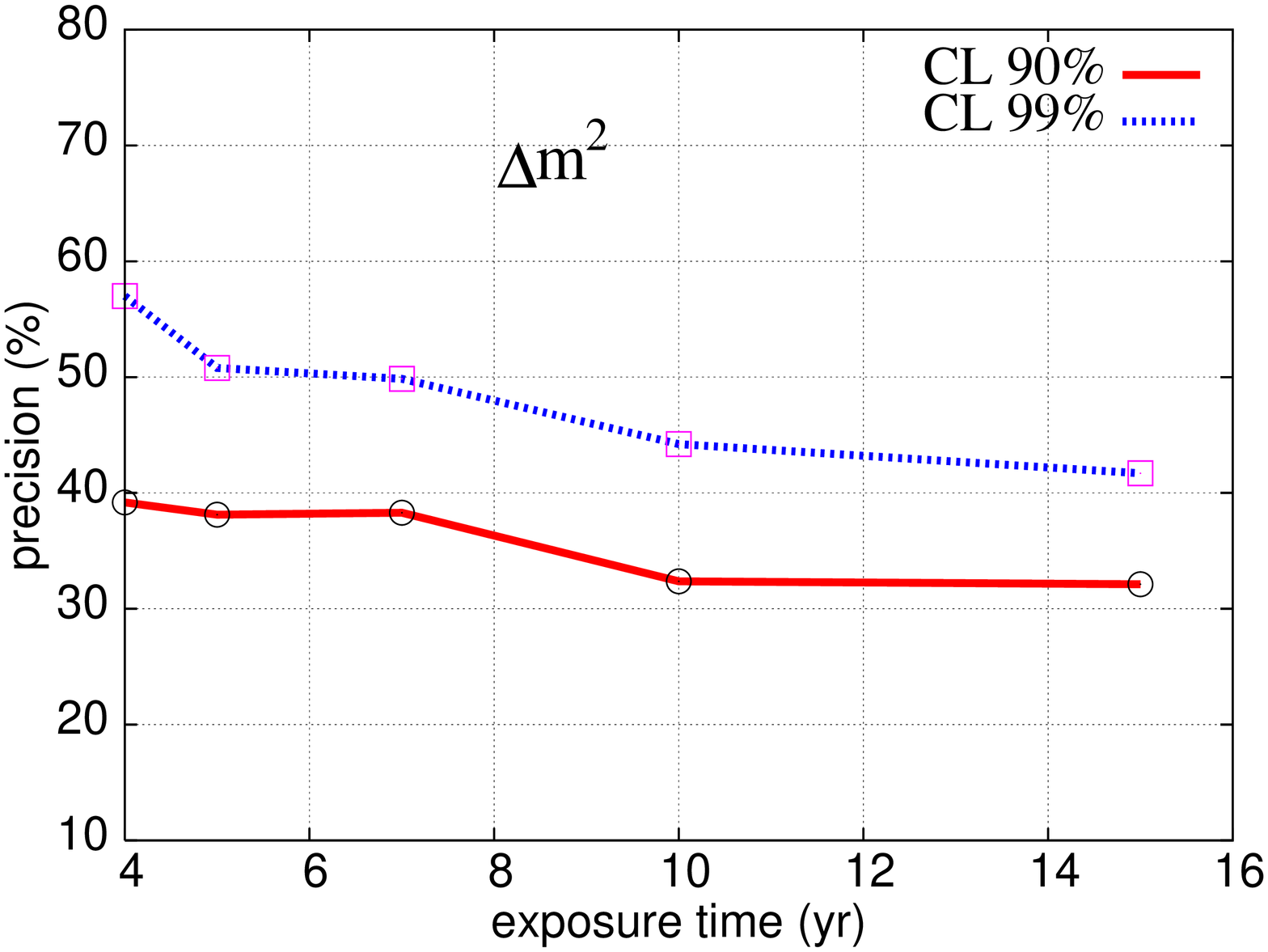}
  \includegraphics[width=6.cm,angle=270]{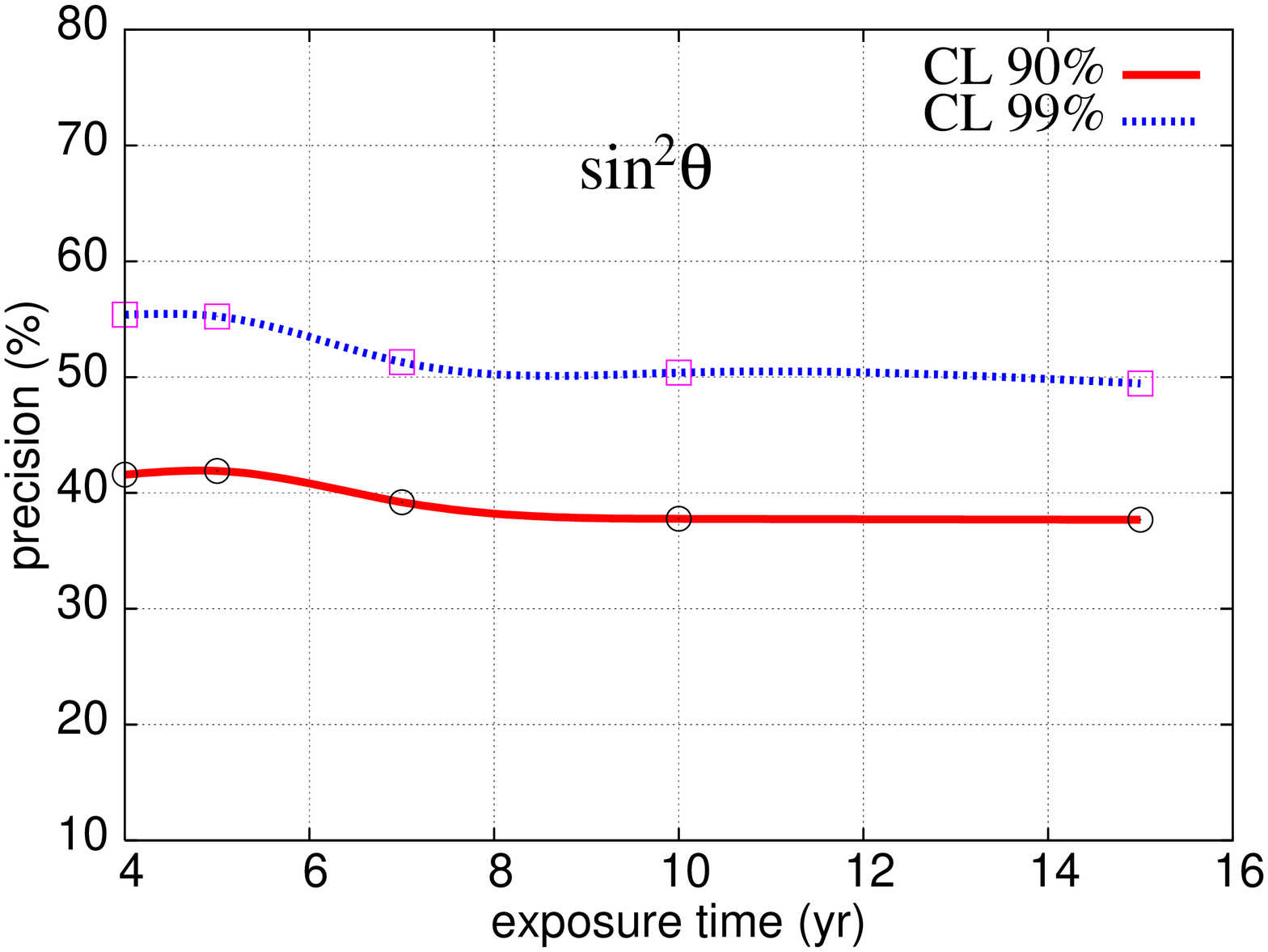} 
\caption{\sf  The variation of the precision 
 of $\Delta m_{}^2$ (left) 
and $\sin^2\theta_{}$ (right) 
with time of exposure
for 50kTon ICAL with  FC
events for the input of $\Delta m^2=2.3\times 10^{-3}$eV$^2$.
}
\label{f:precision}
\end{figure*}

{\bf FC analysis:}
The variation of precision of $\sin^2\theta$ and $\Delta m^2$
with different years of exposure is shown separately in fig.~\ref{f:precision}
at 90\% and 99\% CL. 
It is seen that the precision falls very slowly beyond ten years 
and that can be a useful observation for the future experiment.

\begin{figure*}[htb]
  \includegraphics[width=6.cm,angle=270]{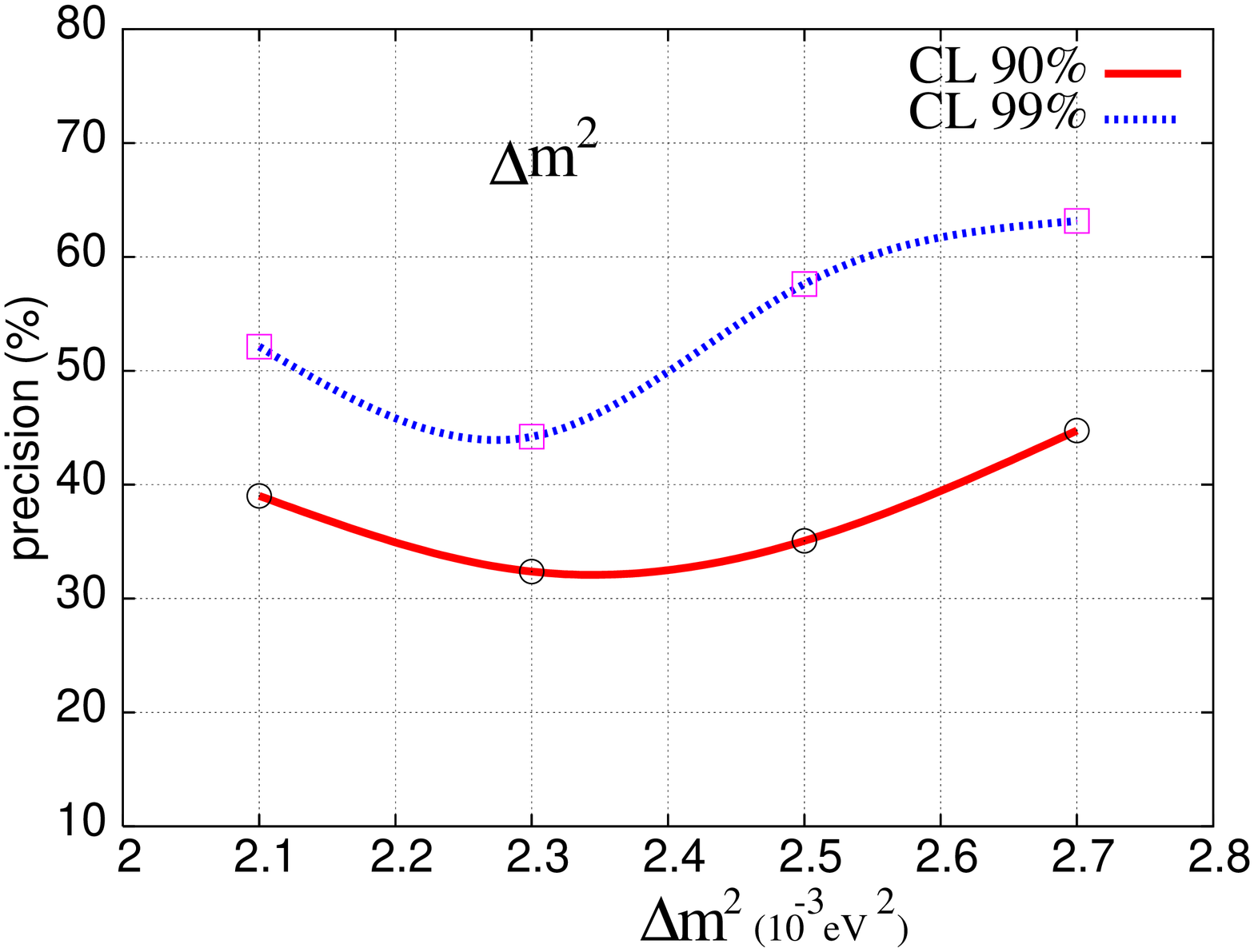}
  \includegraphics[width=6.cm,angle=270]{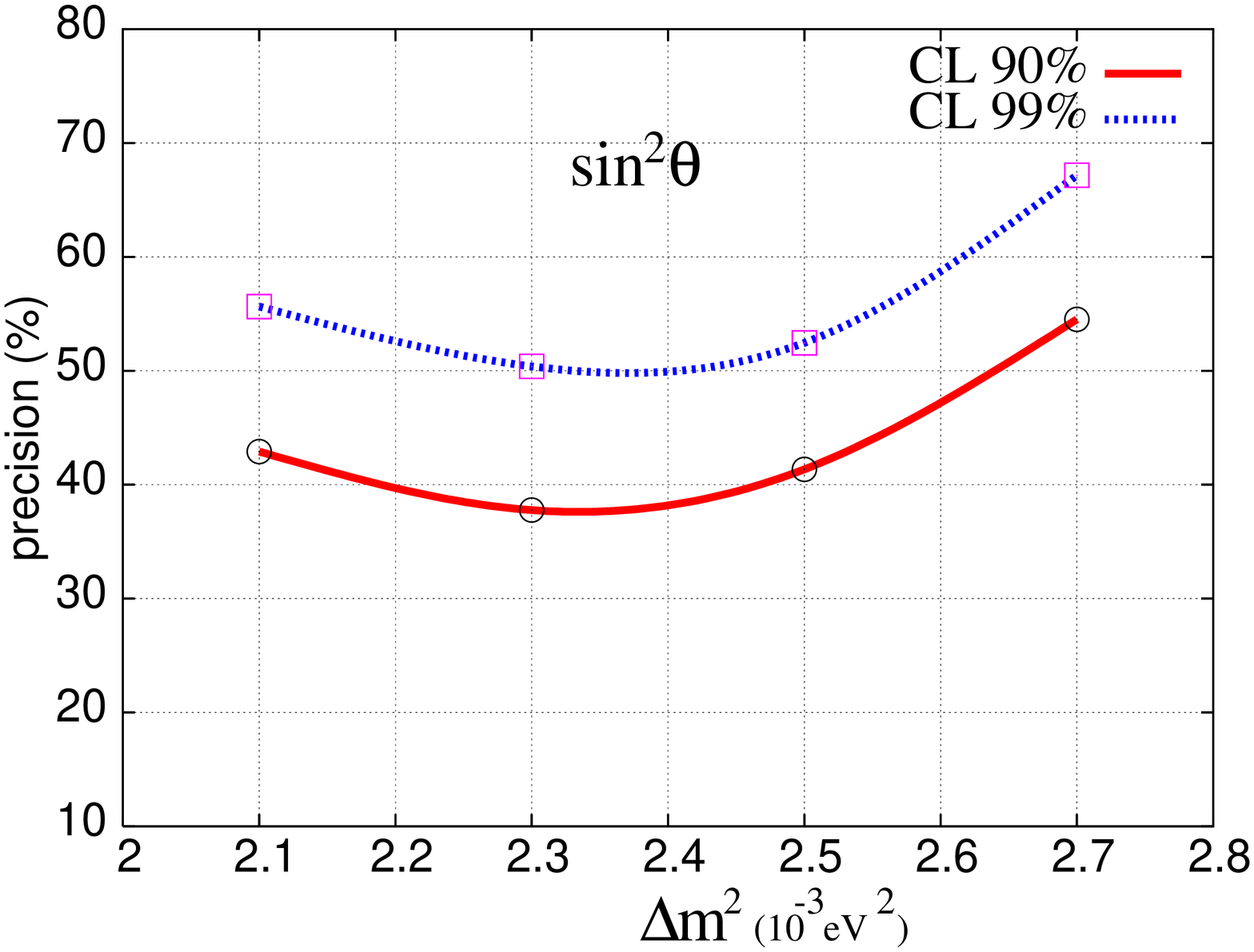}
\caption{\sf 
The variation of the precision  of $\Delta m_{}^2$ (left)
and $\sin^2\theta_{}$ (right)
with the input value of $\Delta m^2$
for 50kTon ICAL with  10 years FC
events.
}
\label{f:precision_delm}
\end{figure*}

It is further observed that the precision gradually becomes worse when we increase 
the value of $\Delta m^2$ from 2.3$\times 10^{-3}$eV$^2$.  This is 
demonstrated in  fig. \ref{f:precision_delm}.

The reason behind this is the following. The position of the dip 
in the up/down distribution  shifts
towards larger values of $L/E$ with the decrease of the value of $\Delta m^2$.  
The flux increases rapidly with decrease of energy 
and the statistics becomes gradually high  at larger
$L/E$. 

 However, we comment that there is a competition between statistics and $L/E$
 resolution. We see that at low value of $\Delta m^2$, say
 2.1 $\times 10^{-3}$eV$^2$, the precision worsens compared to that at
 2.3 $\times 10^{-3}$eV$^2$.  This can be improved if we choose
more stringent cuts with good $L/E$ resolution for  larger values of 
$L/E$.
So one has to optimize
between the requirement of statistics and $L/E$-resolution, which depends
mainly on the range of interest of $\Delta m^2_{}$.  

{\bf FC+PC analysis:}
After the inclusion of  the PC events 
the results are very similar to that obtained from FC events and 
hence are not shown separately.

\section{Discussions}

Simulation studies for atmospheric neutrinos at the proposed 
Iron Calorimeter (ICAL) detector at INO have been made with 
a goal to determine the level of precision which may be achieved. 
The oscillated atmospheric neutrino events for a known set of values of 
oscillation parameters are generated with the event generator 
code  (NUANCE) and the simulated signals in the detector are obtained
through the detector simulation code (GEANT) that uses the NUANCE output 
as its inputs.  A $\chi^2$ analysis of the results obtained 
from this simulated GEANT output 
data, 
properly chosen using appropriate constraints (``cuts"), 
is performed for the precision studies.

There is, however, scope for improvement of these studies. 

\begin{itemize}

\item The present analysis is performed only with the simulated muon 
signals neglecting the hadrons. The 
estimation of neutrino energy $E$ and $L/E$ is expected to improve 
by the inclusion of hadrons.  
As hadrons mainly produce showers instead of 
well defined tracks,  the method of using tracklength  
is not effective 
to extract energy information (or directional information) 
from such hadron showers. A new 
methodology is to be developed for this purpose and 
this work of incorporating hadrons in the analysis is in progress.

\item  
{ A full three flavor analysis can address issues like matter effect,
mass hierarchy and the deviation from maximality of the atmospheric
mixing.} 


\item For the analysis of PC events the curvature of the track is 
used for calculation of energy. In the high energy regime the tracks 
have no or negligible curvature inside the detector volume and hence 
such PC events could not be considered in the present analysis. 
\end{itemize}

Moreover, we find the resolutions are energy dependent and significantly
different for neutrinos and anti-neutrinos. One therefore expects to 
obtain more precise best fit values with CL contours in parameter 
space further shrunk, if one uses  multiple resolution functions, 
instead of one as used here. 
In doing so, the whole $L-E$ plane is divided into multiple 
small segments (mesh) and separate resolution functions
are obtained for each such segment of the mesh which is then used 
for the purpose of analysis. 


{\large{\bf {Acknowledgements}}}

The authors  thank Gobinda Majumder, Subhendu Rakshit, 
Sunanda Banerjee, Subhashis Chattopadhyay and Naba K Mondal for their 
help and valuable suggestions at different phases of the work.


\begin{thebibliography}{99}

\bibitem{Fukuda:1998mi} 
  Y.~Fukuda {\it et al.}  [Super-Kamiokande Collaboration],
  Phys.\ Rev.\ Lett.\  {\bf 81}, 1562 (1998)
  [arXiv:hep-ex/9807003].


\bibitem{Eidelman:2004wy}
  S.~Eidelman {\it et al.}  [Particle Data Group],
  Phys.\ Lett.\  B {\bf 592}, 1 (2004).

\bibitem{Pontecorvo:1957cp}
  B.~Pontecorvo,
  Sov.\ Phys.\ JETP {\bf 6}, 429 (1957).



\bibitem{Maki:1962mu}
  Z.~Maki, M.~Nakagawa and S.~Sakata,
  Prog.\ Theor.\ Phys.\  {\bf 28}, 870 (1962).


\bibitem{Schwetz:2006dh}
  T.~Schwetz,
  Phys.\ Scripta {\bf T127}, 1 (2006)
  [arXiv:hep-ph/0606060].

\bibitem{Apollonio:1999ae}
  M.~Apollonio {\it et al.}  [CHOOZ Collaboration],
  Phys.\ Lett.\ B {\bf 466}, 415 (1999)
  [arXiv:hep-ex/9907037].

\bibitem{Bandyopadhyay:2004da}
  A.~Bandyopadhyay, S.~Choubey, S.~Goswami, S.~T.~Petcov and D.~P.~Roy,
  Phys.\ Lett.\ B {\bf 608}, 115 (2005)
  [arXiv:hep-ph/0406328].

\bibitem{Ashie:2004mr}
  Y.~Ashie {\it et al.}  [Super-Kamiokande Collaboration],
  Phys.\ Rev.\ Lett.\  {\bf 93}, 101801 (2004)
  [arXiv:hep-ex/0404034].

\bibitem{Zois:2004ns}
  M.~G.~Zois,
FERMILAB-MASTERS-2004-06.
\bibitem{Michael:2006rx}
 D.~G.~Michael {\it et al.}  [MINOS Collaboration],
  Phys.\ Rev.\ Lett.\  {\bf 97}, 191801 (2006)
  [arXiv:hep-ex/0607088].

\bibitem{Yamada:2006hi}
  Y.~Yamada  [T2K Collaboration],
  Nucl.\ Phys.\ Proc.\ Suppl.\  {\bf 155}, 207 (2006).

\bibitem{Kisiel:2005ti}
  J.~Kisiel  [ICARUS Collaboration],
  Acta Phys.\ Polon.\ B {\bf 36}, 3227 (2005).

\bibitem{Rubbia:1998rc}
  A.~Rubbia  [ICARUS-CERN-Milano Collaboration],
  Nucl.\ Phys.\ Proc.\ Suppl.\  {\bf 66}, 436 (1998).

\bibitem{Ray:2006ke}
  R.~Ray,
  Nucl.\ Phys.\ Proc.\ Suppl.\  {\bf 154}, 179 (2006).

\bibitem{Harris:2005yb}
  D.~A.~Harris  [MINOS and NOvA Collaborations],
  Nucl.\ Phys.\ Proc.\ Suppl.\  {\bf 149}, 150 (2005).

\bibitem{Horton-Smith:2006yh}
  G.~Horton-Smith  [Double Chooz Collaboration],
  AIP Conf.\ Proc.\  {\bf 805}, 142 (2006).

\bibitem{Motta:2006jd}
 D.~Motta  [Double Chooz Collaboration],
  Acta Phys.\ Polon.\  B {\bf 37}, 2027 (2006).

\bibitem{Jung:1999jq}
  C.~K.~Jung,
  arXiv:hep-ex/0005046.



\bibitem{Back:2004qi}
  H.~Back {\it et al.},
  arXiv:hep-ex/0412016.

\bibitem{Itow:2001ee}
Y.~Itow {\it et al.},
  arXiv:hep-ex/0106019.
%
\bibitem{Nakamura:2003hk}
  K.~Nakamura,
  Int.\ J.\ Mod.\ Phys.\ A {\bf 18}, 4053 (2003).

\bibitem{Cocco:2000yp}
 A.~G.~Cocco  [OPERA Collaboration],
  Nucl.\ Phys.\ Proc.\ Suppl.\  {\bf 85}, 125 (2000).
%
\bibitem{Gustavino:2006rc}
 C.~Gustavino  [OPERA Collaboration],
  J.\ Phys.\ Conf.\ Ser.\  {\bf 39}, 326 (2006).
%
\bibitem{Di Capua:2005bd}
  F.~Di Capua  [OPERA Collaboration],
  PoS {\bf HEP2005}, 177 (2006).

\bibitem{Athar:2006yb}
  M.~S.~Athar {\it et al.}  [INO Collaboration],
   INO-2006-01.

\bibitem{Indumathi:2004kd}
  D.~Indumathi and M.~V.~N.~Murthy,
  Phys.\ Rev.\ D {\bf 71}, 013001 (2005)
  [arXiv:hep-ph/0407336].
%
\bibitem{Gandhi:2004bj}
 R.~Gandhi, P.~Ghoshal, S.~Goswami, P.~Mehta and S.~Uma Sankar,
  Phys.\ Rev.\  D {\bf 73}  053001 (2006)
  [arXiv:hep-ph/0411252].

\bibitem{Petcov:2005rv}
  S.~T.~Petcov and T.~Schwetz,
  Nucl.\ Phys.\ B {\bf 740}, 1 (2006)
  [arXiv:hep-ph/0511277].
%
\bibitem{Samanta:2006sj}
  A.~Samanta,
  arXiv:hep-ph/0610196.

\bibitem{Choubey:2005zy}
  S.~Choubey and P.~Roy,
  Phys.\ Rev.\  D {\bf 73}, 013006 (2006)
  [arXiv:hep-ph/0509197].

\bibitem{Indumathi:2006gr}
 D.~Indumathi, M.~V.~N.~Murthy, G.~Rajasekaran and N.~Sinha,
  Phys.\ Rev.\  D {\bf 74}, 053004 (2006)
  [arXiv:hep-ph/0603264].



%
%
%
%
%
%
\bibitem{Sanuki:2000wh}
 T.~Sanuki {\it et al.},
  Astrophys.\ J.\  {\bf 545}, 1135 (2000)
  [arXiv:astro-ph/0002481].
%
\bibitem{Maeno:2000qx}
  T.~Maeno {\it et al.}  [BESS Collaboration],
  Astropart.\ Phys.\  {\bf 16}, 121 (2001)
  [arXiv:astro-ph/0010381].


\bibitem{Alcaraz:2000vp}
  J.~Alcaraz {\it et al.}  [AMS Collaboration],
  Phys.\ Lett.\  B {\bf 490}, 27 (2000).



\bibitem{Honda:2004yz}
  M.~Honda, T.~Kajita, K.~Kasahara and S.~Midorikawa,
  Phys.\ Rev.\ D {\bf 70}, 043008 (2004)
  [arXiv:astro-ph/0404457].

\bibitem{Casper:2002sd}
  D.~Casper,
  Nucl.\ Phys.\ Proc.\ Suppl.\  {\bf 112}, 161 (2002)
  [arXiv:hep-ph/0208030].


                                                                                



%
%
\bibitem{geant}
GEANT – Detector Simulation and Simulation Tool, CERN Program Library Long Write-up W5013, March 1994, 
http://wwwasd.web.cern.ch/wwwasd/cernlib/version.html

\bibitem{mirrorL}
P. Picchi and F. Pietropaolo, ICGF RAP. INT.
344/1997, Torino 1997 (CERN preprint SCAN-9710037).


\bibitem{Ashie:2005ik}
  Y.~Ashie {\it et al.}  [Super-Kamiokande Collaboration],
  Phys.\ Rev.\  D {\bf 71}, 112005 (2005)
  [arXiv:hep-ex/0501064].

\end{thebibliography}
\end{document}